\documentclass[twocolumn,prl,amsmath,amssymb,aps,showpacs,superscriptaddress]{revtex4-1}

\usepackage{graphicx}
\usepackage{dcolumn}
\usepackage{bm}

\newcommand{\helseven}{\ifmmode {{}^7_{\Lambda}{\rm He}} \else
  {${}^7_{\Lambda}{\rm He}$}\fi}
\newcommand{\helfour}{\ifmmode {{}^4_{\Lambda}{\rm He}} \else
  {${}^4_{\Lambda}{\rm He}$}\fi}
\newcommand{\hlfour}{\ifmmode {{}^4_{\Lambda}{\rm H}} \else
  {${}^4_{\Lambda}{\rm H}$}\fi}
\newcommand{\belseven}{\ifmmode {{}^{7}_{\Lambda}{\rm Be}} \else
  {${}^{7}_{\Lambda}{\rm Be}$}\fi}
\newcommand{\belten}{\ifmmode {{}^{10}_{\Lambda}{\rm Be}} \else
  {${}^{10}_{\Lambda}{\rm Be}$}\fi}
\newcommand{\lilseven}{\ifmmode {{}^{7}_{\Lambda}{\rm Li}} \else
  {${}^{7}_{\Lambda}{\rm Li}$}\fi}
\newcommand{\eekp}{\ifmmode (e,e^{\prime}K^+) \else
  $(e,e^{\prime}K^+)$\fi}

\begin{document}

\preprint{APS/001-He7L}

\title{Observation of the {\helseven} hypernucleus by the {\eekp} reaction}

\newcommand*{\TOHOKU}{Graduate School of Science, Tohoku University, Sendai, Miyagi 980-8578, Japan}
\newcommand*{\HAMPTON}{Department of Physics, Hampton University, Virginia 23668, USA}
\newcommand*{\FIU}{Department of Physics, Florida International University, Miami, Florida 33199, USA}
\newcommand*{\YAMAGATA}{Faculty of Science, Yamagata University, Yamagata 990-8560, Japan}
\newcommand*{\KEK}{Institute of Particle and Nuclear Studies, KEK, Tsukuba, Ibaraki 305-0801, Japan}
\newcommand*{\RIKEN}{Institute for Physical and Chemical Research (RIKEN), Wako, Saitama 351-0198, Japan}
\newcommand*{\HOU}{Department of Physics, University of Houston, Houston, Texas 77204, USA}
\newcommand*{\JLAB}{Thomas Jefferson National Accelerator Facility, Newport News, Virginia 23606, USA}
\newcommand*{\YERPHY}{Yerevan Physics Institute, Yerevan 0036, Armenia}
\newcommand*{\LANZHOU}{Nuclear Physics Institute, Lanzhou University, Lanzhou, Gansu 730000, China}
\newcommand*{\ZAGREB}{Department of Physics \& Department of Applied Physics, University of Zagreb, HR-41001 Zagreb, Croatia}
\newcommand*{\NCSU}{Department of Physics, North Carolina A\&T State University, Greesboro, North Carolina 27411, USA}
\newcommand*{\LATECH}{Department of Physics, Louisiana Tech University, Ruston, Lousiana 71272, USA}
\newcommand*{\JMU}{Department of Physics, James Madison University, Harrisonburg, Virginia 22807, USA}
\newcommand*{\NCW}{Department of Physics, University of North Carolina Wilmington, Wilmington, NC 28403, USA}
\newcommand*{\DUKE}{Triangle Universities Nuclear Laboratory, Duke University, Durham, North Carolina 27708, USA}
\newcommand*{\MARYLAND}{Department of Physics, University of Maryland, College Park, Maryland 20742, USA}
\newcommand*{\SUNO}{Department of Physics, Southern University at New Orleans, New Orleans, LA 70126, USA}
\newcommand*{\CALSTATE}{Physics and Astronomy Department, California State University, Sacramento CA 95819, USA}

\author{S.~N.~Nakamura}
\affiliation{\TOHOKU}
\author{A.~Matsumura}
\affiliation{\TOHOKU}
\author{Y.~Okayasu}
\affiliation{\TOHOKU}
\author{T.~Seva}
\affiliation{\ZAGREB}
\author{V.~M.~Rodriguez} 
\affiliation{\HOU}
\author{P.~Baturin}
\affiliation{\FIU}
\author{L.~Yuan}
\affiliation{\HAMPTON}
\author{A.~Acha}
\affiliation{\FIU}
\author{A.~Ahmidouch}
\affiliation{\NCSU}
\author{D.~Androic}
\affiliation{\ZAGREB}
\author{A.~Asaturyan} 
\affiliation{\YERPHY}
\author{R.~Asaturyan}
\affiliation{\YERPHY}
\author{O.~K.~Baker}
\affiliation{\HAMPTON}
\author{F.~Benmokhtar}
\affiliation{\MARYLAND}
\author{P.~Bosted}
\affiliation{\JLAB}
\author{R.~Carlini}
\affiliation{\JLAB}
\author{C.~Chen}
\affiliation{\HAMPTON}
\author{M.~Christy}
\affiliation{\HAMPTON}
\author{L.~Cole}
\affiliation{\HAMPTON}
\author{S.~Danagoulian}
\affiliation{\NCSU}
\author{A.~Daniel}
\affiliation{\HOU}
\author{V.~Dharmawardane}
\affiliation{\JLAB}
\author{K.~Egiyan}
\affiliation{\YERPHY}
\author{M.~Elaasar}
\affiliation{\SUNO}
\author{R.~Ent}
\affiliation{\JLAB}
\affiliation{\HAMPTON}
\author{H.~Fenker}
\affiliation{\JLAB}
\author{Y.~Fujii}
\affiliation{\TOHOKU}
\author{M.~Furic}
\affiliation{\ZAGREB}
\author{L.~Gan}
\affiliation{\NCW}
\author{D.~Gaskell}
\affiliation{\JLAB}
\author{A.~Gasparian}
\affiliation{\NCSU}
\author{E.~F.~Gibson}
\affiliation{\CALSTATE}
\author{T.~Gogami}
\affiliation{\TOHOKU}
\author{P.~Gueye}
\affiliation{\HAMPTON}
\author{Y.~Han}
\affiliation{\HAMPTON}
\author{O.~Hashimoto}
\affiliation{\TOHOKU}
\author{E.~Hiyama}
\affiliation{\RIKEN}
\author{D.~Honda}
\affiliation{\TOHOKU}
\author{T.~Horn}
\affiliation{\MARYLAND}
\author{B.~Hu}
\affiliation{\LANZHOU}
\author{Ed~V.~Hungerford}
\affiliation{\HOU}
\author{C.~Jayalath}
\affiliation{\HAMPTON}
\author{M.~Jones}
\affiliation{\JLAB}
\author{K.~Johnston} 
\affiliation{\LATECH}
\author{N.~Kalantarians}
\affiliation{\HOU}
\author{H.~Kanda}
\affiliation{\TOHOKU}
\author{M.~Kaneta}
\affiliation{\TOHOKU}
\author{F.~Kato}
\affiliation{\TOHOKU}
\author{S.~Kato}
\affiliation{\YAMAGATA}
\author{D.~Kawama}
\affiliation{\TOHOKU}
\author{C.~Keppel}
\affiliation{\HAMPTON}
\affiliation{\JLAB}
\author{K.~J.~Lan}
\affiliation{\HOU}
\author{W.~Luo}         
\affiliation{\LANZHOU}
\author{D.~Mack}
\affiliation{\JLAB}
\author{K.~Maeda}
\affiliation{\TOHOKU}
\author{S.~Malace}
\affiliation{\HAMPTON}
\author{A.~Margaryan}
\affiliation{\YERPHY}
\author{G.~Marikyan}
\affiliation{\YERPHY}
\author{P.~Markowitz}
\affiliation{\FIU}
\author{T.~Maruta}
\affiliation{\TOHOKU}
\author{N.~Maruyama} 
\affiliation{\TOHOKU}
\author{T.~Miyoshi}
\affiliation{\HOU}
\author{A.~Mkrtchyan} 
\affiliation{\YERPHY}
\author{H.~Mkrtchyan}
\affiliation{\YERPHY}
\author{S.~Nagao}
\affiliation{\TOHOKU}
\author{T.~Navasardyan}
\affiliation{\YERPHY}
\author{G.~Niculescu}
\affiliation{\JMU}
\author{M.-I.~Niculescu}
\affiliation{\JMU}
\author{H.~Nomura}
\affiliation{\TOHOKU}
\author{K.~Nonaka}
\affiliation{\TOHOKU}
\author{A.~Ohtani}
\affiliation{\TOHOKU}
\author{M.~Oyamada}
\affiliation{\TOHOKU}
\author{N.~Perez}
\affiliation{\FIU}
\author{T.~Petkovic}
\affiliation{\ZAGREB}
\author{S.~Randeniya}
\affiliation{\HOU}
\author{J.~Reinhold}
\affiliation{\FIU}
\author{J.~Roche}
\affiliation{\JLAB}
\author{Y.~Sato}
\affiliation{\KEK}
\author{E.~K.~Segbefia}
\affiliation{\HAMPTON}
\author{N.~Simicevic}
\affiliation{\LATECH}
\author{G.~Smith}
\affiliation{\JLAB}
\author{Y.~Song}
\affiliation{\LANZHOU}
\author{M.~Sumihama}
\affiliation{\TOHOKU}
\author{V.~Tadevosyan}
\affiliation{\YERPHY}
\author{T.~Takahashi}
\affiliation{\TOHOKU}
\author{L.~Tang}
\affiliation{\HAMPTON}
\affiliation{\JLAB}
\author{K.~Tsukada}
\affiliation{\TOHOKU}
\author{V.~Tvaskis}
\affiliation{\HAMPTON}
\author{W.~Vulcan}
\affiliation{\JLAB}
\author{S.~Wells}
\affiliation{\LATECH}
\author{S.~A.~Wood}
\affiliation{\JLAB}
\author{C.~Yan}
\affiliation{\JLAB}
\author{S.~Zhamkochyan} 
\affiliation{\YERPHY}

\collaboration{HKS (JLab E01-011) Collaboration}

\date{\today}

\begin{abstract}
An experiment with a newly developed high-resolution kaon spectrometer (HKS) 
and a scattered electron spectrometer with a novel configuration was
performed in Hall C at Jefferson Lab (JLab).
The ground state of a neutron-rich hypernucleus, {\helseven}, was
observed for the first time with the {\eekp} reaction with an energy
resolution of $\sim$0.6~MeV.  This resolution is the best reported to
date for hypernuclear reaction spectroscopy.
The {\helseven} binding energy supplies the last missing information of the $A=7, T=1$ hypernuclear iso-triplet,
providing a new input for the charge symmetry breaking (CSB) effect of $\Lambda N$ potential.
\end{abstract}

\maketitle

Our world consists of electrons and nucleons (protons and neutrons) which 
are composed of up and down valence quarks. 
In a hyperon, one of these original quarks is replaced by a strange quark. 
A hypernucleus contains a hyperon implanted as an impurity within the nuclear medium. 
The lightest hyperon is the $\Lambda$ particle (up+down+strange, isospin 0). 
Precise information about the mass and excitation
energies of hypernuclei allows one to infer the underlying hyperon-nucleon (YN) interaction 
which is relevant to the discussion of high density nuclear matter such as neutron stars.
Precise nucleon-nucleon (NN) potentials have been derived from the rich dataset of
nucleon scattering experiments as well as from
the masses and excitation energies of nuclei.  In contrast, hyperon-nucleon (YN)
scattering experiments are technically difficult and data is very limited.  Therefore,
hypernuclear spectroscopy is a
more realistic method to study the YN interaction.

The study of hypernuclei seeks to extend our knowledge of the nuclear force and baryon-baryon 
forces in general. While the strange quark is heavier than up and down quarks, it is
light enough to be treated in the framework of SU(3)$_{\rm flavor}$ symmetry, a natural extension of
isospin symmetry for nucleons. Understanding of baryon-baryon forces based on SU(3)$_{\rm flavor}$
symmetry is important to bridge between phenomenologically well studied nuclear force
models and the underlying degrees of freedom of the strong interaction as described by
Quantum Chromodynamics (QCD).

Since a single $\Lambda$ inside a nucleus is not subject to the Pauli Exclusion Principle, it can occupy any accessible
shell, including the deeply bound s-shell in the heaviest nuclei. The $\Lambda$ decays 
with a relatively long lifetime, even in a nucleus ($\sim$200~ps) \cite{Outa, Qiu}, and thus the widths of hypernuclear
energy levels are typically less than a few 100 keV.  This fact makes the spectroscopic study of these systems
 possible. The $\Lambda$ can also probe the interior structure of the host nucleus. Furthermore, one can
search for possible modifications of the composition and structure of deeply bound baryons \cite{Dover, E13}.


After the first observation of a $\Lambda$ hypernucleus more than a half century ago with an emulsion \cite{FirstHY}, meson 
beams such as $K^-$ and $\pi^+$ have been widely used to obtain spectroscopic information via missing mass analysis in the
${}^AZ(K^-,\pi^-){}^A_{\Lambda}Z$ and ${}^AZ(\pi^+,K^+){}^A_{\Lambda}Z$ reactions. 
In both of these reactions, $\Lambda$ hyperons are produced off neutrons, which precludes 
the use of the elementary reaction channel for an accurate mass calibration. 
Together with the inherently limited quality of these secondary meson beams, 
the accuracy of absolute mass determinations has been limited to a resolution of no better than 1.5~MeV.

The ${}^A Z{\eekp}{}^A_{\Lambda}(Z\!-\!1)$ reaction produces strangeness by s-$\overline{\rm s}$ 
pair-production, similar to the $(\pi^+,K^+)$ reaction. 
An interesting feature of the {\eekp} reaction is that it converts a proton 
to a $\Lambda$
enabling us to calibrate the absolute missing mass scale 
by using the $p{\eekp}\Lambda, \Sigma^0$ reactions with the well known masses of the $\Lambda$ and $\Sigma^0$ hyperons.  
Furthermore, the {\eekp} reaction can produce new species of hypernuclei and 
thus the charge dependence of hypernuclei can be studied by comparing {\eekp} hypernuclear spectroscopy 
to already known iso-multiplet partners.
As well as the above unique features, {\eekp} hypernuclear reaction
spectroscopy has the potential for good (sub-MeV) energy resolution
due to the availability of primary electron beams with lower energy spread
than available for secondary meson beams.

We report here the first clear observation of the ground state of 
{\helseven} through the $^{7}\mbox{Li}{\eekp}{}^7_{\Lambda}\mbox{He}$ reaction. 
Although {\helseven} has been observed in emulsion experiments \cite{Juric}, 
only a total of 11 events are known; furthermore, the measured masses for these events are spread out widely, 
which lead to speculation that long lived isomeric states \cite{PNIE1,DALITZ,PNIE2} were 
observed together with the ground state. 
Therefore, no ground state mass has been quoted in the literature.

{\helseven} is the missing member of the $A=7$, $T=1$ isospin triplet, 
the other two being $\lilseven^*$ and {\belseven}.
The three core nuclei ${}^6{\rm He}$, ${}^6{\rm Li}^*$, and ${}^6{\rm Be}$ have in common an $\alpha$-core, 
surrounded by a halo nucleon pair, $nn$, $pn$, and $pp$, respectively. 
Likewise, the bound $\Lambda$ wave function is predicted to reach far beyond the $\alpha$-core and thus have a significant overlap with the halo nucleon pair \cite{Hiyama}.
In particular, {\helseven} plays a key role in the study of the halo structure of neutron-rich hypernuclei since 
it has a core of the lightest bound neutron-halo nucleus ${}^6{\rm He}$.
 
As suggested by Hiyama {\it et al.}, this iso-triplet is the perfect
testing ground to study the Charge Symmetry Breaking (CSB) effect 
in the $\Lambda$N potential.
The 
binding energies of the iso-triplet were recently computed using a four-body cluster model with  the CSB effect \cite{Hiyama}. 
A $\Lambda$N CSB potential
was phenomenologically introduced to explain the binding energy difference of the $A=4$
iso-doublet ($T=1/2$) hypernuclei {\hlfour} and {\helfour}. 
The difference
$B_\Lambda({}^4_\Lambda{\rm He})-B_\Lambda({}^4_\Lambda{\rm H}) = +0.35\pm 0.06$ MeV
is unexpectedly large even  after corrections due to the Coulomb interaction.



The experimental challenge of {\eekp} hypernuclear reaction spectroscopy originates from 
the small hypernuclear production cross section and high backgrounds. The cross section 
of the {\eekp} reaction is $\leq 100$ nb/sr which is two-three
orders of magnitude smaller than that of hadronic production. 
Furthermore, the {\eekp} reaction requires two spectrometers for a
coincidence between scattered electrons and kaons.
These experimental difficulties 
result in lower hypernuclear yields and poorer
signal-to-noise ratios than meson reactions.

The pilot experiment E89-009 (HNSS), performed at Jefferson Lab (JLab)
in 2000, demonstrated
the principle of {\eekp} hypernuclear spectroscopy \cite{Miyoshi,Lulin}.
The experiment showed that
${}^{12}{\rm C}{\eekp}{}^{12}_\Lambda{\rm B}$ spectroscopy with sub-MeV resolution 
is possible with the high quality electron beam at JLab, but it also showed
that improvements were possible to fully exploit the potential of hypernuclear study  
by electro-production.
The $10^{-3}$ momentum resolution and small solid angle of the kaon spectrometer (SOS)
limited the resolution and hypernuclear yield.  
In E89-009, zero-degree electrons were measured to maximize the
virtual photon yield, but this also reduced the signal-to-noise ratio
thus limiting the beam current and target thickness that could be used.

The natural extension to the E89-009 experiment is the E01-011
experiment (HKS) performed in JLab's Hall C in 2005
\cite{E01-011prop} from which the spectrum discussed here was obtained.
\begin{figure}[h]
\includegraphics[width=8cm]{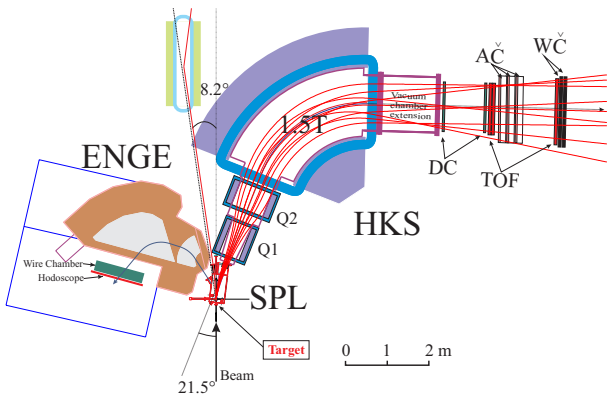}
\caption{\label{fig:Setup} Schematic figure of the E01-011 experimental setup. A newly constructed HKS (High resolution Kaon Spectrometer) and an
ENGE-type split pole spectrometer (ENGE) which was also used in E89-009 were used as the kaon and scattered electron spectrometers. The ENGE was vertically 
tilted to suppress background originating from Bremsstrahlung and M$\phi$ller scattering.}
\end{figure}
In the E01-011 experiment, a high resolution ($\Delta p/p \sim 2 \times 10^{-4}$) kaon spectrometer (HKS) was constructed and an existing
electron spectrometer was optimized 
to improve the resolution and hypernuclear yield over the E89-009 experiment.
The improvements allowed the system to handle 180 times higher luminosity (4.5 times thicker target and 40 times more intense electron beam) with 100 times smaller 
electron background rate in the electron spectrometer.

Figure~\ref{fig:Setup} shows a layout of the experimental setup.
The target was placed in a dipole magnet (SPL) which separated the oppositely charged particles at small forward angles. 
The $K^+$s were measured by the HKS which has a central momentum of $P_K = 1.2~{\rm GeV}/c$ and a 16~msr solid angle when used with the SPL magnet.
The scattered electrons (central momentum $P_{e^{\prime}} = 0.35~{\rm GeV}/c$) were measured 
by the ENGE-type split-pole spectrometer which was vertically tilted by 
8~degrees from the dispersion plane and shifted vertically by
an amount to suppress electron backgrounds originating from Bremsstrahlung and M$\phi$ller scattering which have very sharp
forward distributions (tilt method).
The electron beam energy was set at $E_e = 1.851~{\rm GeV}$ giving
a virtual photon energy of about 1.5~GeV ($ \simeq E_e - c\:P_{e^{\prime}}$). 
The typical beam current for the lithium target was 25~$\mu$A.
Details of the design of the experiment will be explained elsewhere \cite{HKS-NIM}.
Since the beam energy from CEBAF at JLab was known with an accuracy of $1 \times 10^{-4} \sim 180$~keV,
measurements of the  momentum vectors of $K^+$ and $e^{\prime}$ at the target were sufficient to obtain the missing mass of the hypernuclei.
The positions and angles of the scattered kaons and electrons were measured
at the focal planes of the HKS and ENGE spectrometers.  These focal plane
quantities were converted to target momentum vectors using backward
transfer matrices of the spectrometers.
The initial transfer matrices were generated by using a GEANT4 Monte
Carlo simulation with three-dimensional magnetic field maps 
obtained by field measurements and finite element calculations by Opera-3D (TOSCA). 
The backward transfer matrices were obtained from these initial
transfer matrices and tuned using calibration data such as the 
sieve slit data which constrains the angular parts of the matrices,
and the $\Lambda$ and $\Sigma^0$ peaks from the $p{\eekp}\Lambda, \Sigma^0$
reaction with protons in a ${\rm CH}_2$ target, constraining the
the momentum parts of the matrices.

\begin{figure}[htb]
\includegraphics[width=8cm]{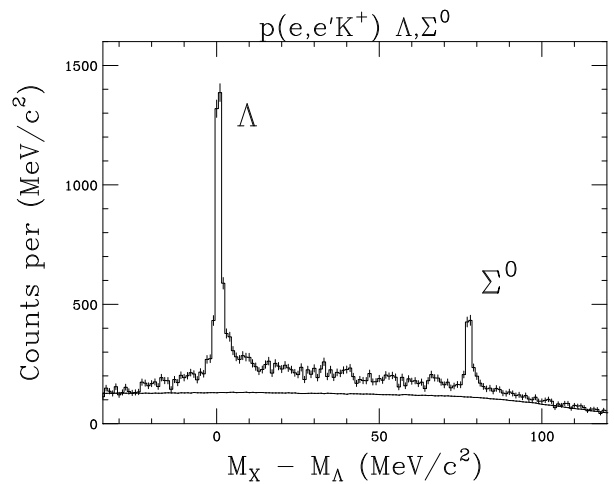}
\caption{\label{fig:LambdaSigma} Missing mass spectrum of the $p{\eekp}\Lambda/\Sigma^0$ reaction. Mass of $\Lambda$ particle was subtracted. 
$\Lambda$, $\Sigma^0$ peaks were used to calibrate 
the absolute missing mass scale. The line shows the accidental background estimated by the mixed events analysis method.}
\end{figure}

Figure~\ref{fig:LambdaSigma} shows the missing mass spectrum from scattering
off a ${\rm CH}_2$ target with clear peaks corresponding to
$\Lambda$ and $\Sigma^0$ hyperon production off of protons and an underlying
background from quasi-free hyperon production on carbon and 
accidental coincidences between $e^{\prime}$s and $K^+$s.
The background shape from the quasi-free hyperon production on carbon was measured by using a
${}^{12}{\rm C}$ target and 
the accidental background shape was obtained by randomly selecting uncorrelated $e^{\prime}$s and $K^+$s (mixed events
analysis). These events were sampled from real data with an off-time gate in the coincidence timing of $e^\prime$ and $K^+$.
Therefore, it is ensured that the mixed events and the background have the same momentum distributions.
The background shape of the ${\rm CH}_2$ target data was almost completely described by the above contributions and thus 
we can conclude that the mixed event analysis technique can be safely applied to estimate the accidental background shapes in our analysis.

The $\Lambda$ and $\Sigma^0$ peak positions were used for missing mass calibration and the backward transfer matrix tunes. 
The tuned and calibrated matrices gave the peak positions in
table~\ref{tab:L-S0}. The missing mass scale was calibrated for these
hyperons within a 100~keV uncertainty.
The widths of hyperon peaks are worse than the expected sub-MeV resolution for hypernuclei because hyperons are much lighter than hypernuclei and kinematic broadening due to finite angular resolution of spectrometers contributed more significantly to the energy resolution.

\begin{table}[b]
\caption{\label{tab:L-S0}%
$\Lambda$ and $\Sigma^0$ masses: $M_Y$ (PDG values \cite{PDG}) and $M_X$ fitted values of E01-011 data. Unit of values is MeV/$c^2$.
}
\begin{ruledtabular}
\begin{tabular}{cccc}
\textrm{Hyperon}& $M_Y$& 
\textrm{$M_X - M_Y$}&
\textrm{Width (FWHM)}\\
\colrule
$\Lambda$ & 1115.683 & $0.09 \pm 0.02$ & $1.94 \pm 0.45$\\
$\Sigma^0$ & 1192.642 & $0.05 \pm 0.03$ & $1.87 \pm 0.56$
\end{tabular}
\end{ruledtabular}
\end{table}

\begin{figure}[thb]
\includegraphics[width=5.9cm]{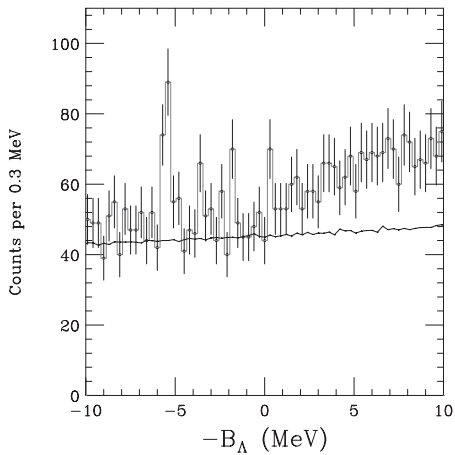}\vspace*{-3mm}
\caption{\label{fig:RawHe7L} Binding energy spectrum obtained by the 
${}^7{\rm Li}{\eekp}{\helseven}$ reaction. The line shows the accidental background estimated by the mixed events analysis method.
The events in the unbound region ($-B_\Lambda > 0$) originate from quasi-free $\Lambda$ production.}
\end{figure}

In the E01-011 experiment, a natural Li target of 189~mg/cm$^2$ (${}^7{\rm Li}$ abundance 92.4\%), was used as a target.
The measured missing mass was converted to binding energy using:
\[-B_\Lambda = M({\helseven}) - (M_\Lambda + M({}^6{\rm He})),\]
and plotted in figure~\ref{fig:RawHe7L}.
A ${}^6{\rm He}$ mass of $5605.537~{\rm MeV}/c^2$ was obtained from the
reported mass excess \cite{Audi}.
\begin{figure}[thb]
\includegraphics[width=5.9cm]{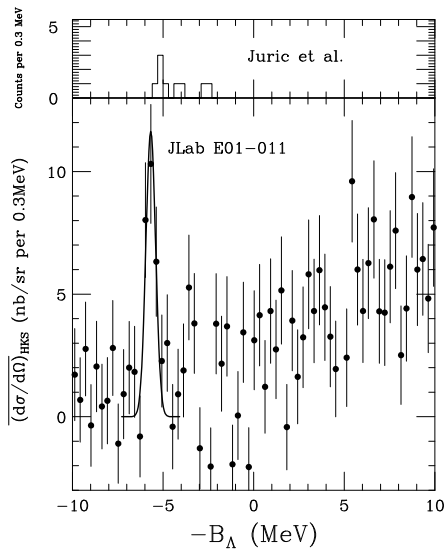}\vspace*{-3mm}
\caption{\label{fig:He7L-HKS} Binding energy spectra of {\helseven} measured by the emulsion experiment \cite{Juric} (top) 
and the JLab E01-011 (HKS) experiment after background subtraction and acceptance corrections (bottom).}
\end{figure}
The accidental coincidence events in figure~\ref{fig:RawHe7L} were estimated by using the mixed events technique. 
After subtraction of the accidental background and correction of the spectrometers' acceptances and detector efficiencies, 
the number of counts was converted to the differential cross section averaged over the acceptance of the
HKS ($1.05 < P_K < 1.35~{\rm GeV}/c$, $1^\circ < \theta_K < 13^\circ$).
Since the virtual photon is almost real ($Q^2 \sim 0.01~\mbox{GeV}^2/c^2, 
W \sim 1.9~\mbox{GeV}$),
the {\eekp} differential cross section was converted to the 
differential cross section for virtual photons using 
the virtual photon flux ($\Gamma$) as:
\[
 \frac{d\sigma}{d\Omega_K} = \frac{1}{\Gamma} \frac{d \sigma}{dE_{e'} d\Omega_{e'} d\Omega_{K}}.  \]

The virtual photon flux integrated over the ENGE acceptance 
($0.24 < P_{e^{\prime}} < 0.44~{\rm GeV}/c$, $\Delta \Omega_{e'} = 5.6$~msr) was 
$4.8 \times 10^{-6}$ virtual photons per electron.

Figure~\ref{fig:He7L-HKS} shows the {\helseven} spectra measured by the emulsion experiment (top) \cite{Juric} and 
by JLab E01-011 (bottom).
The E01-011 spectrum shows a clear peak which corresponds to the {\helseven} ground state ($1/2^+$).

The number of events in the peak ($S$ for $-6.15< -B_\Lambda < -4.65$~MeV) and 
the number of background events under the peak ($N$) were counted in the raw spectrum (Fig.~\ref{fig:RawHe7L}) to obtain
the peak significance:
\[ S/\sqrt{S+N} = 97/\sqrt{316} = 5.5,\] 
which is consistent with the statistical error of the cross section obtained from the fit of the acceptance corrected spectrum to be shown.
Different choices for the background in the region of the peak 
result in lower peak significance.  However, 
the peak retains a high likelihood of being real and 
its existence is quite solid.

There exists some structure between the ground state and the threshold ($B_\Lambda = 0$), but the statistics are not enough 
to discuss in detail.
The systematic error of the binding energy due to the tuning processes of the transfer matrices was estimated 
by applying the analysis procedures to dummy data generated by a full Monte Carlo simulation 
with arbitrarily chosen hypernuclear masses and various signal-to-noise ratios ($S/N$).
The simulated data were analyzed using the same software as the real data and the arbitrarily chosen hypernuclear masses 
were hidden from the analysis group. The difference between the inputs to the simulation and the analysis results were treated as 
the systematic error due to the matrix tuning processes.
The estimated systematic error depends on $S/N$. For major peaks ($S/N>0.3$), the error was less than 100~keV, but for 
poor $S/N$ peaks ($S/N < 0.3$), the error is as large as 400~keV.
Other sources of systematic error on the binding energy are uncertainties in the kinematic parameters such as the absolute electron beam energy and the central momenta of the $K^+$ and $e^{\prime}$. These contributions were studied carefully and estimated to be less than 150~keV.
The systematic errors on the cross section were estimated with the same method and combined
with the beam current uncertainty.

The ground state peak of {\helseven} was fitted with a Gaussian.  The binding energy and virtual photon cross section obtained were:
\begin{eqnarray*}
-B_\Lambda = -5.68 \pm 0.03 (\rm stat.) \pm 0.25 (\rm sys.) \: {\rm MeV},\\
\left(\overline{\frac{d\sigma}{d\Omega}}\right)_{\rm HKS}  =  26 \pm 5.1 {\rm (stat.)} \pm 9.9 {\rm (sys.)}\: {\rm nb/sr},  
\end{eqnarray*}
with a  width of $0.63 \pm 0.12$~MeV (FWHM).

The E01-011 experiment successfully observed the {\helseven} ground state with sufficient statistics.
The emulsion data show a cluster with a broad tail (Fig.~\ref{fig:He7L-HKS} top) and the binding energy was not obtained \cite{Juric}.
It was hypothesized that the cluster corresponded to the ground state and 
that the broad tail originated from the decay of isomeric states of {\helseven} 
but this was not experimentally confirmed \cite{PNIE1,DALITZ,PNIE2}.
The E01-011 data are consistent with the interpretation that the cluster of emulsion data corresponds to the ground state.

The binding energies of the {\lilseven} and {\belseven} ground states were measured by emulsion \cite{Juric} but 
the ground state of {\lilseven} is the $T=0$ state ($B_\Lambda({\lilseven}, T=0) = 5.58 \pm 0.03$~MeV) \cite{Juric}.
Therefore, the energy spacing information from the $\gamma$-ray measurement, $Ex(T=1,1/2^+) = 3.88$~MeV \cite{Tamura} and 
the excitation energy of ${}^6{\rm Li}^* (T=1) =3.56$~MeV were used to calculate the binding energy of $\lilseven^* (T=1)$ state.

The binding energies of the $A=7,~T=1$ iso-triplet hypernuclei, {\helseven}, $\lilseven^*$, {\belseven}, now experimentally 
measured are shown in table~\ref{tab:T=1A=7}.

\begin{table}[hbt]
\caption{\label{tab:T=1A=7}%
Binding energies of the $A=7, T=1$ iso-triplet $\Lambda$ hypernuclei. 
E01-011 errors are statistical and systematic.
}
\begin{ruledtabular}
\begin{tabular}{cccc}
            & {\helseven} \textrm{(E01-011)}& $\lilseven^*$ \cite{Juric,Tamura}& {\belseven} \cite{Juric}\\
\colrule
$B_\Lambda$ (MeV) & $5.68 \pm 0.03 \pm 0.25$ & $5.26 \pm 0.03$ & $5.16 \pm 0.08$\\
\end{tabular}
\end{ruledtabular}
\end{table}


The binding energies of the $A = 7$ hypernuclear iso-triplet can provide useful information about the CSB effect of the $\Lambda$N potential
by comparing to the results of an $\alpha N N \Lambda$ four-body cluster calculation
as has been done with the $A=4$ ($NNN\Lambda$; $^4_\Lambda$H, $^4_\Lambda$He) and $A=10$ ($\alpha \alpha N \Lambda$; $^{10}_\Lambda$Be, $^{10}_\Lambda$B) 
hypernuclei \cite{HiyamaA=10, Zhang}.

The {\eekp} hypernuclear spectroscopy technique was established at JLab by the present E01-011 experiment in Hall C 
and an independent experiment (E94-107) performed in Hall A \cite{N16L,Li9L}. 
The experimental efforts have continued 
to improve with the JLab E05-115 experiment \cite{E05-115prop} and a recently initiated program at the upgraded MAMI-C, Mainz University \cite{KaoS,KaoS2}.


As the binding energy difference between {\hlfour} and {\helfour} hypernuclei is the starting point of CSB discussions, 
new measurements with recent experimental techniques are necessary.
Binding energy measurements of {\hlfour} are planned at both JLab and MAMI-C by using the ${}^4{\rm He}{\eekp}{\hlfour}$ reaction and there are plans to use the
newly proposed decay $\pi$ spectroscopy of hyperfragments technique \cite{Tang} and to do a hypernuclear $\gamma$-ray experiment of {\helfour} at J-PARC \cite{E13}. 
More precise data on {\helseven} and {\belten} with a better control of the systematic errors from JLab E05-115 and the planned experiments on $A=4$ hypernuclei will provide definitive experimental information to determine the CSB terms in the $\Lambda N$ potential.

\vspace{3mm}

We acknowledge continuous support and encouragement from the staff of the Jefferson Lab physics and accelerator divisions.
The hypernuclear programs at JLab Hall-C are supported by Japan-MEXT Grant-in-aid 
for Scientific Research (16GS0201, 15684005, 12002001, 08239102, 09304028, 09554007, 11440070, 15204014), 
Japan-US collaborative research program, Core-to-core program (21002) and 
strategic young researcher overseas visits program 
for accelerating brain circulation (R2201) by JSPS,
US-DOE contracts (DE-AC05-84ER40150, DE-FG02-99ER41065, DE-FG02-97ER41047, DE-AC02-06CH11357, DE-FG02-00ER41110, DE-AC02-98-CH10886) and US-NSF (013815,0758095).


\end{document}